\documentclass[10pt,journal,twocolumn]{IEEEtran}
\newif\ifCLASSOPTIONromanappendices \CLASSOPTIONromanappendicestrue
\usepackage{a0size}
\usepackage{amssymb}
\usepackage{multicol}
\usepackage[english]{babel}
\usepackage{epsfig}
\usepackage{bm}
\usepackage{amsfonts,color,amsthm,amsmath, lscape,graphics}
\usepackage{url}
\usepackage{algorithm}
\usepackage{algpseudocode}

\usepackage[font=footnotesize]{caption}
\usepackage{subcaption}
\usepackage{epstopdf}
\usepackage{algorithm}
\usepackage{algpseudocode}
\usepackage{cite}
\usepackage{relsize}

\DeclareFontFamily{U}{matha}{\hyphenchar\font45}
\DeclareFontShape{U}{matha}{m}{n}{
      <5> <6> <7> <8> <9> <10> gen * matha
      <10.95> matha10 <12> <14.4> <17.28> <20.74> <24.88> matha12
      }{}
\DeclareSymbolFont{matha}{U}{matha}{m}{n}
\DeclareMathSymbol{\odiv}         {2}{matha}{"63}

\usepackage{fix2col}

\DeclareMathOperator*{\Minimize}{minimize}

\renewcommand{\figurename}{Fig.}
\addto\captionsenglish{\renewcommand{\figurename}{Fig.}}

\newcommand{\bb}{\mathbf{b}}

\newcommand{\bh}{\mathbf{h}}

\newcommand{\bv}{\mathbf{v}}

\newcommand{\bU}{\mathbf{U}}
\newcommand{\bI}{\mathbf{I}}
\newcommand{\bw}{\mathbf{w}}
\newcommand{\bx}{\mathbf{x}}

\newcommand{\bs}{\mathbf{s}}

\newcommand{\by}{\mathbf{y}}
\newcommand{\bz}{\mathbf{z}}
\newcommand{\bA}{\mathbf{A}}

\newcommand{\bC}{\mathbf{C}}

\newcommand{\bH}{\mathbf{H}}

\newcommand{\bM}{\mathbf{M}}

\newcommand{\bq}{\mathbf{q}}

\newcommand{\bW}{\mathbf{W}}

\renewcommand{\frac}{\dfrac}

\definecolor{myOrange}{rgb}{1,0.5,0}
\definecolor{myGreen}{rgb}{0,0.5,0}

\newcommand{\changeF}[1]{{\color{black}#1}}

\newcommand{\changeN}[1]{{\color{black}#1}}

\begin{document}
\title{Learning Progressive Distributed Compression Strategies from Local Channel State Information 
}

\author{{Foad~Sohrabi},~\IEEEmembership{Member,~IEEE,}
        Tao~Jiang,~\IEEEmembership{Graduate Student Member,~IEEE,}
        and~Wei~Yu,~\IEEEmembership{Fellow,~IEEE}
\thanks{The authors are with The Edward S.\ Rogers Sr.\ Department of
Electrical and Computer Engineering, University of Toronto, Toronto, ON M5S
3G4, Canada (e-mails: \{fsohrabi, tjiang, weiyu\}@ece.utoronto.ca). This work
is supported in part by the Natural Sciences and Engineering Research Council
(NSERC) via the Canada Research Chairs program and in part by Huawei Technologies Canada.
}
}
\maketitle
\begin{abstract}

This paper proposes a deep learning framework to design distributed compression strategies in which distributed agents need to compress high-dimensional observations of a source, then send the compressed bits via bandwidth limited links to a fusion center for source reconstruction. Further, we require the compression strategy to be progressive so that it can adapt to the varying link bandwidths between the agents and the fusion center. Moreover, to ensure scalability, we investigate strategies that depend only on the local channel state information (CSI) at each agent. Toward this end, we use a data-driven approach in which the progressive linear combination and uniform quantization strategy at each agent are trained as a function of its local CSI. To deal with the challenges of modeling the quantization operations (which always produce zero gradients in the training of neural networks), we propose a novel approach of exploiting the statistics of the batch training data to set the dynamic ranges of the uniform quantizers. Numerically, we show that the proposed distributed estimation strategy designed with only local CSI can significantly reduce the signaling overhead and can achieve a lower mean-squared error distortion for source reconstruction than state-of-the-art designs that require global CSI at comparable overall communication cost.

\end{abstract}


\section{Introduction}
\label{sec:intro}

Distributed estimation is a fundamental problem in many signal processing applications. For example, in a wireless sensor network consisting of multiple spatially distributed sensors connected to a fusion center via bandwidth limited links, a common task is to reconstruct the source at the fusion center---with applications ranging from remote sensing, environmental monitoring, to surveillance, localization, tracking, edge AI inference, etc., \cite{Wong2002SPMag,Akyildiz2002Survey,TomLuo2006Mag,9134426}.  As another example, in a cloud radio access network (C-RAN), also known as cell-free massive multiple-input multiple-output (MIMO) network, a central processor needs to decode the information transmitted by a user based on the signals received at geographically distributed base-stations \cite{Liang2015CRAN,Wei2017BookChapter}. A common characteristic in many of these applications is that the observations at the distributed agents are typically high dimensional, yet the communication links between the agents and the central fusion center (also known as the fronthaul) are bandwidth limited. A key consideration in the design of distributed estimation schemes is therefore how to optimally reduce the dimension of the observations and to compress them at the distributed agents to satisfy the bandwidth constraints.

Toward this end, this paper considers a general model in which each agent observes a noisy version of a source signal through a linear channel; the agents are connected to a fusion center via bandwidth limited links; and the ultimate goal is to reconstruct the source at the fusion center. To reduce the communications between the agents and the fusion center, each agent applies a linear compression matrix to its high-dimensional observation, followed by a bank of per-dimension uniform quantizers. Upon receiving the compressed and quantized signals from all agents, the fusion center employs a linear estimator to recover the original source.

The optimal distributed compression and estimation scheme at the agents should be a function of the channel state information (CSI) from the source to all the agents.  However, requiring global CSI presents scalability issues in terms of signaling overhead between the agents and the fusion center. This is because the agents are distributed in space, and a strategy requiring global CSI would need to collect all the CSI centrally. In this paper, we investigate the performance of the compression and quantization strategies designed solely based on the \emph{local CSI} at each agent. Specifically, this paper shows that by adopting a data-driven approach, it is possible to \emph{learn} the optimized local compression strategy at each agent based on its own CSI. Because of the reduced signaling overhead, such an approach can actually achieve a lower distortion when the channels need to be frequently re-estimated from time to time, as compared to a centralized design requiring global CSI, at comparable overall communication cost. 

In many practical wireless distributed systems, the qualities of the links between the agents and the fusion center are often time varying, i.e., the number of bits each agent can transmit to the fusion center may vary from time slot to time slot. \emph{Progressive compression} is a concept that gives each agent the capability of dynamically adjusting its compression rate according to the instantaneous conditions of the fronthaul links, while achieving the optimal rate-distortion frontier.  Progressive coding is desirable for applications in which signal reconstruction is time sensitive. For example, in surveillance applications, it is beneficial to reconstruct a low-resolution image immediately (e.g., for detecting a potential threat), even if the communication links between the sensors and the fusion center temporarily experience deep fading \cite{Gunduz2019MLProgressive}. \changeN{This paper shows that with} minor modifications, the proposed learning framework can be adapted to make the compression strategy \emph{distributed} and \emph{progressive}, while based only on \emph{local} CSI.

\subsection{Main Contributions}
Most of the existing works in distributed estimation literature assume that the global CSI is available at the fusion center for the purpose of compression matrices design. However, for this assumption to hold in practical large-scale networks, extensive communication overhead between the agents and the fusion center is needed. This is because, in such schemes, the fusion center needs to first collect the high-dimensional observation matrices (i.e., CSI) from all the agents, and after designing the compression schemes, it must send the designed compression matrices back to the agents. In order to reduce such communication overhead, it is desirable to devise compression strategies at the agents based on the locally available CSI. But the design of local compression strategies for distributed systems based on the conventional optimization-based methods is quite challenging because it is not straightforward to define a local optimization objective function at each agent. Accordingly, the existing compression strategies based on local CSI typically focus on the heuristical objective of maximizing the local compression performance for each agent, e.g., through principal component analysis (PCA). In this paper, we show that such PCA-based local compression strategies are not optimal, and much more efficient local compression and quantization schemes can be designed via deep learning.   

Moreover, as mentioned earlier, we are interested in designing progressive compression strategies, to deal with scenarios in which the capacities of the links between the agents and the fusion center are time varying. However, the associated constraints for developing a progressive transmission scheme are difficult to satisfy in most of the existing model-based methods. In this paper, we show that the challenges of designing a progressive compression strategy for distributed wireless systems can also be tackled using a data-driven deep learning approach.

The proposed deep learning framework in this paper consists of a fully-connected deep neural network (DNN) at each agent for designing the corresponding compression and quantization schemes and a linear minimum mean square error (LMMSE) estimator at the fusion center for recovering the source signal. In particular, the agent-side DNNs take the local CSI  as input and outputs the design of the compression matrices employed by the agents to reduce the dimension of the observed signals. After dimension reduction, the observation signals are uniformly quantized and subsequently transmitted to the fusion center for signal estimation. Finally, the fusion center applies an LMMSE estimator in order to recover the source signal based on the quantized signals received from all the agents.

The training of the proposed DNN, which involves an intermediate quantization layer, is not a trivial task. This is because a quantization layer gives vanishing gradients, which inhibit the training of the network parameters prior to that layer through the conventional back-propagation algorithm. To get around with this issue, a straight-through estimator is widely used in machine learning literature, in which the quantization operation is approximated with another smooth and differentiable function \cite{hinton2012videos,bengio2013,chung2016}. But, for some application settings, e.g. \cite{Foad2019Asilomar}, we observe that training of a DNN with quantization hidden layers using the straight-through estimator is still cumbersome. This paper proposes an alternative novel technique for training a DNN with an intermediate uniform quantization layer by modeling the quantization error as uniformly distributed noise, the range of which is a function of the quantization dynamic range and the number of quantization bits. However, the setting of the quantization dynamic range is still a challenging task, because in the training phase the design of the compression matrices is always changing due to the update of the DNN parameters, thus affecting the dynamic range. To deal with this so-called \emph{internal covariate shift} phenomenon \cite{Ioffe2015}, this paper proposes to set the dynamic range as a scaled version of the standard deviation of the signals which can be empirically computed from the batch data. Moreover, the scaling factor can be learned by using back-propagation training. In the testing phase, the signals can be quantized according to the uniform quantizer with the dynamic range learned from the training phase. 

Numerical results show that the proposed deep learning framework significantly outperforms the existing PCA-based method which uses local CSI. Further, when compared with compression strategies using global CSI, if the overhead of communicating CSI is accounted for, the proposed deep learning method can actually achieve a lower distortion than the state-of-the-art global CSI-based methods at comparable total signaling cost.

\subsection{Related Work}
One common way to meet the power and bandwidth budget constraints inherent in battery-powered wireless edge sensing devices (i.e., the agents) is to reduce the dimensionality of the observed signals, e.g., \cite{TomLuo2005NP,Zhu2005EVD,Liang2015CRAN,Poor2015Sparse,Zhang2003OptCond,TomLuo2007BCD2,Roy2008BCD1,Fang2008Assignment1,Zhang2019Assignment2,Zhang2021Assignment3}. The paper \cite{TomLuo2005NP} shows that the problem of designing the optimal compression matrices for distributed estimation is in general nonconvex and NP-hard. To tackle this challenging optimization problem, the earlier works on dimensionality reduction in distributed estimation advocate the idea of PCA, in which each agent seeks to optimally compress its observations based on the available local information; see \cite{Zhu2005EVD} in the context of wireless sensor network and \cite{Liang2015CRAN} in the context of multi-antenna C-RAN. Such PCA-based methods mainly use eigenvalue decomposition and subspace projection. Although the PCA-based method leads to the optimal solution for a single agent scenario \cite{Zhang2003OptCond}, for scenarios with multiple agents it is shown in \cite{TomLuo2007BCD2} that the performance of such methods can be significantly improved by using iterative block coordinate descent (BCD). Further, \cite{Roy2008BCD1} numerically investigates the optimality properties of the BCD algorithm, and shows that although for some scenarios (especially those with correlated sources), there is a gap between the performance of the optimal design and the one obtained from the BCD algorithm, this gap is often marginal. It should be mentioned that, in each iteration of the BCD algorithm, the design of the compression matrix at each agent is updated based on an optimal closed-form solution for given compression matrices at the other agents. As a result, in order to perform the steps of the BCD algorithm, the optimizer needs to have access to the CSI of all agents, i.e., global CSI. The follow-up works in this line of research further consider dimension assignment (to each agent) in the design of the compression strategy, under the assumption that the total number of compression dimensions is given, e.g., \cite{Fang2008Assignment1,Zhang2019Assignment2,Zhang2021Assignment3}. Nevertheless, all of these methods are based on the assumption that the global CSI is available for compression design. 

In this paper, we recognize that such global CSI-based compression strategies impose a significant amount of signaling overhead in a distributed wireless system. This is because to design compression strategies based on global CSI, all the CSI needs to be collected at the fusion center, then once the compression matrices are designed, they need to be sent back to the agents. Instead of relying on global CSI, this paper proposes a deep learning framework to design the compression strategies at the agents which only take the local CSI as the input. We numerically show that the proposed local CSI-based method can significantly reduce the signaling overhead of global CSI-based methods \cite{TomLuo2007BCD2,Roy2008BCD1}, while being superior to existing local CSI-based methods such as \cite{Zhu2005EVD,Liang2015CRAN} which are based on local eigenvalue decomposition. 

The aforementioned references \cite{TomLuo2005NP,Zhu2005EVD,Liang2015CRAN,Poor2015Sparse,Zhang2003OptCond,TomLuo2007BCD2,Roy2008BCD1,Fang2008Assignment1,Zhang2019Assignment2,Zhang2021Assignment3} tackle the distributed estimation problem by treating the source signal as a random parameter vector with known distribution. However, motivated by the fact that in some applications obtaining the statistical information of the source may not be feasible, there are other works in the literature (e.g., \cite{Zhang2019DetAWGN,Fang2009DetVQ,Giannakis2006DetNonGaus,Giannakis2006DetAWGN,Fang2010Inhomo} and references therein) that treat the source as a deterministic vector. In this paper, we design the compression matrices at the agents using a data-driven approach, where the empirical distributions of the source signal, channels, and noise are implicitly exploited in the training phase. Further, to set the aggregation matrix at the fusion center, this paper employs the LMMSE estimator which explicitly uses the distribution of the source and the noise.

The distributed compression problem has long been studied in the information theory literature under the names of distributed source coding \cite{1055037,1055508}, or the CEO problem \cite{490552}, for which the focus has been the optimal use of the binning strategies for both lossless and lossy compression. This paper mostly follows the signal processing literature and restricts to linear processing strategies at the agents followed by uniform quantization. The optimization of binning strategies would be considerably more involved.

As already mentioned, this paper seeks to design distributed compression strategies at the agents that are \emph{progressive}, in order to accommodate the dynamic transmission conditions.  This is known as the \emph{successive refinement} problem in the information theory literature. The paper \cite{75242} establishes the necessary and sufficient conditions for successive refinement for the single source coding problem. For example, Gaussian sources are shown to be successively refinable in terms of MSE distortion \cite{75242}.  In \cite{1317111}, successive refinement is also studied for Wyner-Ziv coding, which has applications in distributed source coding for sensor networks \cite{1328091}. For distributed estimation problem, \cite{4475358} shows that the quadratic Gaussian CEO problem can be solved via successive Wyner-Ziv coding. Very recently, \cite{Gunduz2019MLProgressive} proposes a deep learning-based non-linear compression for progressive transmission of image sources in a single-agent scenario. In this paper, we use a deep learning framework to design a distributed linear compression strategy where multiple agents progressively send additional dimensions to the fusion center for estimation purposes. Further, in contrast to \cite{Gunduz2019MLProgressive}, this paper also accounts for the practical quantization step and proposes a novel training procedure to deal with the challenges of training a DNN with intermediate quantization layers.

\subsection{Paper Organization and Notations}
The remainder of this paper is organized as follows. Section~\ref{sec:sys} introduces the system model and the problem formulation for designing a progressive distributed estimation system. Section~\ref{sec:proposed} shows how to design such a distributed compression system using a neural network and further discusses how to train the neural network. Section~\ref{sec:sims} provides simulation results. Finally, conclusions are drawn in Section~\ref{sec:conclusion}.

This paper uses lower-case letters for scalar variables, lower-case bold-face letters for vectors, and upper-case bold-face letters for matrices. Further, we use the superscripts $(\cdot)^\top$ and  $(\cdot)^{-1}$ to denote the transpose and the inverse operations, respectively. The identity matrix with appropriate dimensions is denoted by $\mathbf{I}$; $\mathbb{R}^{m\times n}$ denotes an $m$ by $n$ dimensional real space; $\mathcal{N}(\mathbf{0},\boldsymbol{\Sigma})$ represents the zero-mean circularly symmetric real Gaussian distribution with covariance matrix $\boldsymbol{\Sigma}$; and $\mathcal{U}(a,b)$ represents a uniform distribution on the interval $[a,b]$. The notations $\operatorname{log}_{10}(\cdot)$ and $\mathbb{E} [\cdot] $ represent the decimal logarithm and expectation operators, respectively. Furthermore, $\|\bv\|_2$ indicates the Euclidean norm of a vector $\bv$, $|\mathcal{S}|$ represents the cardinality of a set $\mathcal{S}$, and $\operatorname{vec}(\bM)$ denotes the vectorized representation of a matrix $\bM$. Finally, the hyperbolic tangent activation function is defined as $\operatorname{tanh}(x)\triangleq \tfrac{e^{x}-e^{-x}}{e^{x}+e^{-x}}$.

\section{System Model and Problem Formulation}
\label{sec:sys}

\subsection{Signal Model}
Consider a distributed wireless network consisting of $B$ spatially distributed agents and a fusion center. Each agent makes a noisy observation $\by_i \in \mathbb{R}^{M}$ of the unknown random vector $\bx \in \mathbb{R}^N$, with distribution $f_{\bx}$, according to:
\begin{equation}
\by_i = \bH_i \bx + \bz_i, \quad\quad i = 1,\ldots,B,
\end{equation}
where $\bH_i \in \mathbb{R}^{M \times N}$ is the observation matrix (or the channel matrix) of the $i$-th agent which is assumed to be known at the agent $i$ (or can be estimated, e.g., in the C-RAN application, via pilots), and $\bz_i \sim \mathcal{N}(\mathbf{0},\sigma^2 \bI)$ is the additive white Gaussian noise. Without loss of generality, we assume that the unknown random vector $\bx$ is zero mean. 
Further, we assume that there is no inter-agent communication; each agent needs to send a compressed version of its observation to the fusion center, for the reconstruction of the source $\bx$ at the fusion center. 
Typically, $\bx$ is independent and identically distributed (i.i.d.) over time and needs to be estimated in each time slot, while $\bH_i$ remains stationary over a coherence interval $T$.

This paper assumes the following compression strategy at the distributed agents. Due to the bandwidth limitation on the channel between the agents and the fusion center, the $i$-th agent first applies an appropriate dimension-reducing linear transform to convert the observation vector $\by_i$ into a $K$-dimensional vector with $K \leq M$ as:
\begin{equation}
{\bv}_i = \bW_i \by_i, \quad\quad i = 1,\ldots,B,
\end{equation}
and then sends the quantized version of ${\bv}_i$ to the fusion center as:
\begin{equation}
\tilde{\bv}_i = \mathcal{Q}_i\left( {\bv}_i \right), \quad\quad i = 1,\ldots,B.
\end{equation}
Here $\bW_i \in \mathbb{R}^{K\times M} $ is a full row rank compression matrix employed at the $i$-th agent with compression rate of $\tfrac{K}{M}$. Further, $\mathcal{Q}_i(\cdot)$ denotes a set of scalar $Q$-bit quantizers applied to different dimensions of the reduced dimension signal at the $i$-th agent, i.e., ${\bv}_i$. In this paper, we consider a uniform quantization scheme with a designable dynamic range $q_{i,k}$ for quantizing the $k$-th element of ${\bv}_i$.

Finally, the fusion center collects the quantized signals from all $B$ agents, i.e., $\breve{\bv}= [\tilde{\bv}_1^\top, \tilde{\bv}_2^\top, \ldots,\tilde{\bv}_B^\top]^\top$, and seeks to recover $\bx$ by employing a linear estimator as:
\begin{equation}
\hat{\bx} = \bC \breve{\bv},
\end{equation}
where $\bC \in \mathbb{R}^{N\times BK}$ is the aggregation matrix. 

For simplicity, we assume a symmetric scenario where the dimensions of the observations, the compression rate, and the quantization resolution for all the agents are the same, but the setup can be easily generalized to non-symmetric cases.

\subsection{Non-Progressive Distributed Estimation from Global CSI}
With the above communication models in place, the problem of minimizing the mean squared error (MSE) distortion at the fusion center of a distributed wireless network can be written as:
\begin{subequations}
\label{eq:problem_formulation_noSuc}
\begin{align}
\Minimize_{\{\bW_i, \bq_{i}\}_{\forall i},\bC} &~~~\mathbb{E}\left[ \| \hat{\bx} - \bx \|^2_2 \right]\\
\text{subject to}\hspace{4pt} &~~~\hat{\bx} = \bC \left[\tilde{\bv}_1^\top,\ldots,\tilde{\bv}_B^\top \right]^\top,\\
&~~~ \tilde{\bv}_i= \mathcal{Q}_i\left(\bW_i \bH_i \bx + \bW_i  \bz_i \right),~~ \changeN{\forall i},
\end{align}
\end{subequations}
where $\bq_i \triangleq [q_{i,1},\ldots,q_{i,K}]^\top$ and the expectation in the objective function is over the distribution of $\bx$ and $\{\bz_i\}_{i=1}^B$.

Most of the existing works, e.g., \cite{Roy2008BCD1,TomLuo2007BCD2,Fang2008Assignment1,Zhang2019Assignment2,Zhang2021Assignment3}, solve the distributed estimation problem at the fusion center under the assumption that it has access to the global CSI, i.e., $\{ \bH_i \}_{i=1}^B$. Doing so for large-scale networks requires extensive communication between the agents and the fusion center at each coherence time. In particular, in order to formulate the problem \eqref{eq:problem_formulation_noSuc} at the fusion center, the $i$-th agent must communicate its observation matrix $\mathbf{H}_i$ to the fusion center at the beginning of each coherence interval. Then, the fusion center designs the compression matrices and quantization schemes for all $B$ agents together with the aggregation matrix $\bC$ by solving problem \eqref{eq:problem_formulation_noSuc}. Finally, to deploy the designed compression and quantization schemes at the agents, the fusion center requires to inform agent $i$ about its corresponding designed compression matrix $\bW_i$ and quantization dynamic range vector $\bq_{i}$. 

To account for the signaling overhead in this scheme, we note that each agent needs to transmit $MN$ channel gains to the agent, resulting in $BMN$ \changeN{real-valued} transmission from agents to the fusion center. Further, the fusion center needs to transmit back the design of a $K\times M$ compression matrix and $K$ quantization dynamic ranges to each agent, resulting in $BKM+BK$ \changeN{real-valued} transmission from the fusion center to all $B$ agents. 

To quantify the amount of communications, suppose that within each coherence interval in which the observation matrices remain unchanged, we can perform $T$ different estimation stages. Under this assumption, the overall real-valued transmissions (per coherence interval) required between each agent and the fusion center can be computed for the global CSI-based methods as:
\begin{equation}
    \label{eq:C_global}
        C_\text{global} = \underbrace{MN + MK + K}_{\text{signaling overhead prior to est.\ stages}} + \underbrace{TK}_{\text{signaling in est.\ stages}}.
\end{equation}
This communication burden can be significant. In the next subsection, we show that this communication overhead can be significantly reduced if the agents use the local CSI to design their corresponding compression and quantization schemes.

\begin{figure*}[t]
     \centering
     \includegraphics[width=0.8 \textwidth]{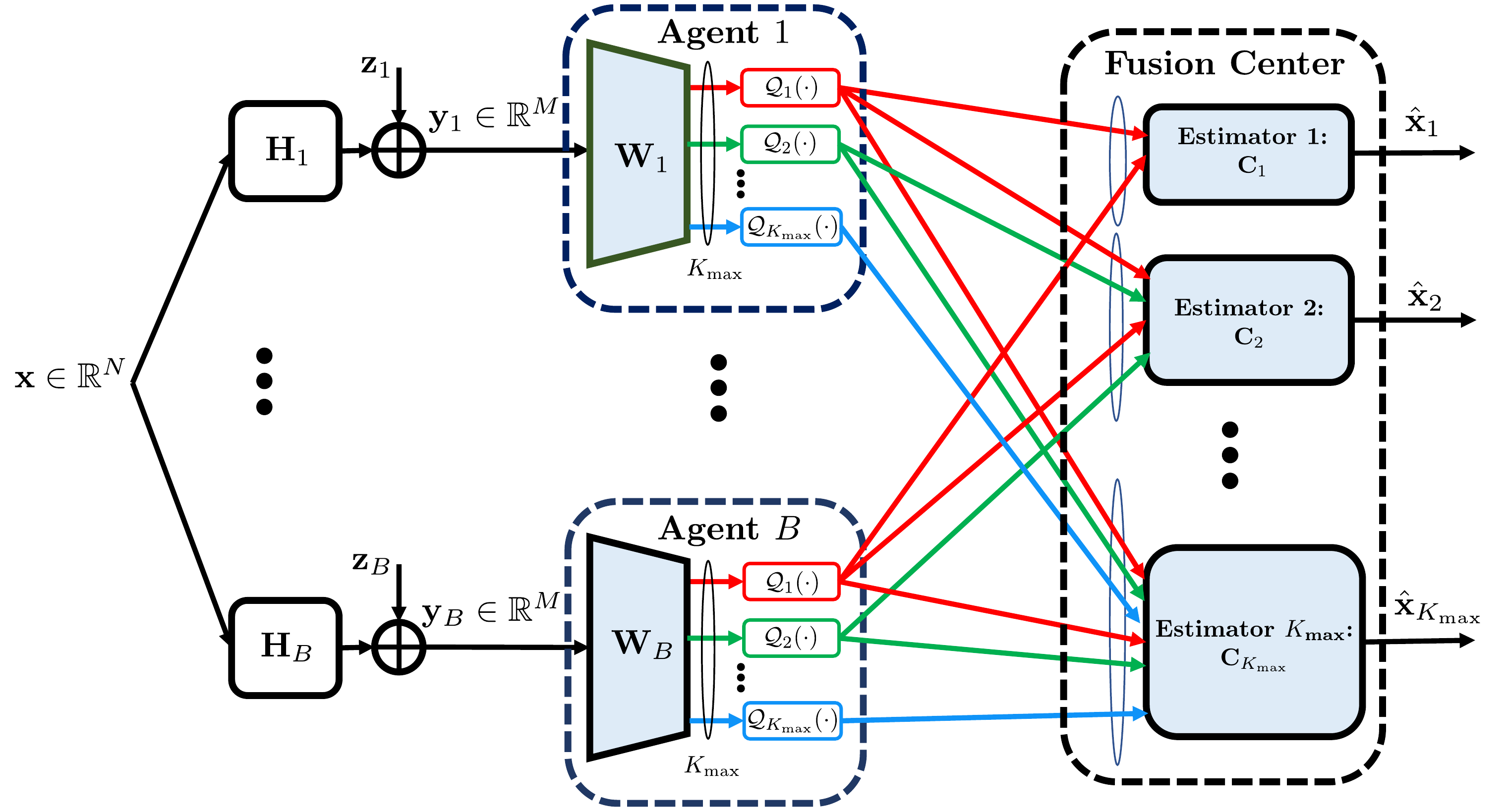}
     \caption{System model for the distributed estimation of a source with progressive compression and reconstruction.}
     \label{fig:sys_model}
\end{figure*} 

\subsection{Non-Progressive Distributed Estimation from Local CSI}
In this paper, we seek to reduce the communications between the agents and the fusion center at the beginning of each coherence interval by solving the problem \eqref{eq:problem_formulation_noSuc} such that the compression matrix at each agent is designed based only on the local channel state information available at that agent, i.e., $\bH_i$, while the aggregation matrix $\bC$ is designed based on the effective observation matrix at the fusion center $\bW_i\bH_i$ and the effective noise covariance $\sigma^2\bW_i\bW_i^\top$. To further reduce the communication overhead between the agents and the fusion center, it is desirable to design the quantization dynamic ranges for each agent based only on \changeN{the statistical  information including the distribution of the observation matrix (denoted by $f_{\bH_i}$), the source distribution, and the noise variance.} Mathematically speaking, we seek to design the optimization variables in \eqref{eq:problem_formulation_noSuc} based on the following mappings:
\begin{subequations}
    \begin{align}
            \mathbf{W}_i &= \mathcal{F}_i(\mathbf{H}_i), \quad \forall i,\\
            \mathbf{q}_i &= \widetilde{\mathcal{F}}_i(f_{\mathbf{H}_i},\changeN{{f_\bx}, \sigma^2}), \quad \forall i,\\
            \mathbf{C} &= \mathcal{G}\left( \left\{\bW_i\bH_i\right\}_{\forall i} ,\left\{\sigma^2\bW_i\bW_i^\top\right\}_{\forall i}\right), 
    \end{align}
\end{subequations}
where functions $\mathcal{F}_i(\cdot)$ and $\widetilde{\mathcal{F}}_i(\cdot)$ respectively determine the compression scheme and the quantization dynamic range at the $i$-th agent, and function $\mathcal{G}_i(\cdot)$ gives the linear estimator used at the fusion center.  

By following such a design strategy, for scenarios in which the observations are high dimensional so that typically $K \ll M$, we can significantly reduce the number of required \changeN{real-valued} transmissions from all agents to the fusion center in the CSI acquisition phase from $BMN$ to $BKN+BK^2$, and further entirely eliminate the need for downlink transmission from the fusion center to the agents, since the compression and quantization schemes are already designed at each agent based on the local information. Following this discussion, the overall real-valued transmissions per coherence interval required between each agent and the fusion center in the local CSI case is given by:
\begin{equation}
        \label{eq:C_local}
        C_\text{local} = \underbrace{KN + K^2}_{\text{signaling overhead prior to est.\ stages}} + \underbrace{TK}_{\text{signaling in est.\ stages}}.
\end{equation}

\changeF{We remark that in practical system implementation, these real-valued scalars all need to be digitized into bits before being transmitted. But since the main focus of this paper is on the fronthaul requirement for the quantization operations in the estimation stages, and since the estimations of $\mathbf{x}$ typically operate at a faster time scale, for the purpose of accounting for the signaling overhead, we make the simplifying assumption that the quantizations in the signaling stage prior to the estimation of $\mathbf{x}$ use fixed and a large number of bits per real value \changeN{in order to ensure perfect reconstruction.}}
\subsection{Progressive Distributed Estimation from Local CSI}

So far, we describe a distributed estimation setting in which each agent can always transmit a fixed number of bits, i.e., $KQ$ bits, to the fusion center for recovering a single vector $\bx$. However, in many practical scenarios, the capacity of the fronthaul links between each agent and the fusion center is time varying. As a result, the number of bits that each agent can transmit to the fusion center may not be known beforehand. It is therefore desirable to design a progressive compression scheme at each agent such that it can adjust its compression rate according to the instantaneous quality of the fronthaul link. Accordingly, in this paper, we investigate a distributed progressive transmission scheme in which agent $i$ progressively compresses the observation $\by_i$ into maximum $K_\text{max}$ dimensions according to:
\begin{equation}
    \tilde{v}_{i,k} = \mathcal{Q}_{i,k}\left( \mathbf{w}_{i,k}^{\top} \by_i\right), \quad\quad k = 1,\ldots,K_\text{max},
\end{equation}
where $\mathcal{Q}_{i,k}(\cdot)$ is a uniform $Q$-bit scalar \changeN{quantizer} with a dynamic range of $q_{i,k}$. Upon receiving the first $k \in \{1,\ldots,K_\text{max}\}$ dimensions from all $B$ agents, i.e., $\tilde{\bv}_{i,[k]} \triangleq [\tilde{v}_{i,1},\ldots,\tilde{v}_{i,k}]^T,~ i=1,\ldots,B$, the fusion center applies the corresponding linear estimator $\bC_k \in \mathbb{R}^{N \times Bk} $ to obtain the following estimate for $\mathbf{x}$:
\begin{equation}
    \hat{\bx}_k = \mathbf{C}_k \left[\tilde{\bv}_{1,[k]}^\top,\ldots,\tilde{\bv}_{B,[k]}^\top \right]^\top,
\end{equation}
with the corresponding MSE distortion of: 
\begin{equation}
    {d}_k = \mathbb{E}\left[ \| \hat{\bx}_k - \bx \|^2_2 \right].
\end{equation}
Thus, if each of the fronthaul links can support $kQ$ bits of transmission, then we achieve a distortion of $d_k$, for each of $k=1,\cdots,K_{\max}$.

For such a system as illustrated in Fig.~\ref{fig:sys_model}, the overall system objective can be thought of as to minimize an aggregated cost function of all possible distortion metrics $\mathcal{D}\left(d_1,\ldots,d_{K_\text{max}}\right)$, while accounting for the different probabilities that the fronthaul links have different capacities. 

With all the aforementioned models and design requirements in place, the overall optimization problem of interest can now be formulated as: 
\begin{subequations}
\label{eq:problem_final}
\begin{align}
\Minimize_{\{\bW_i,\bq_{i}\},\{\bC_k\}} &~\mathcal{D}\left(d_1,\ldots,d_{K_\text{max}}\right)\\
\text{subject to}\hspace{6pt} &~{d}_k = \mathbb{E}\left[ \| \hat{\bx}_k - \bx \|^2_2 \right],\\
&~ \hat{\bx}_k = \mathbf{C}_k \left[\tilde{\bv}_{1,[k]}^\top,\ldots,\tilde{\bv}_{B,[k]}^\top \right]^\top,\\
&~ \tilde{v}_{i,k} = \mathcal{Q}_{i,k}\left( \mathbf{w}_{i,k}^{\top} \by_i\right),\\
&~ \mathbf{W}_i =  \mathcal{F}_i(\mathbf{H}_i),\\
&~ \mathbf{q}_i = \widetilde{\mathcal{F}}_i(f_{\mathbf{H}_i},\changeN{{f_\bx}, \sigma^2}),\\
&~  \mathbf{C}_k = \mathcal{G}\left( \left\{\bW_{i,[k]}\bH_i\right\} ,\left\{\sigma^2\bW_{i,[k]}\bW_{i,[k]}^\top\right\}\right), 
\end{align}
\end{subequations}
where $\bW_{i,[k]} \triangleq [\bw_{i,1},\ldots,\bw_{i,k}]^\top $ and $\bW_i \triangleq \bW_{i,[K_\text{max}]}$.

\begin{figure*}[t]
     \centering
     \includegraphics[width=0.75\textwidth]{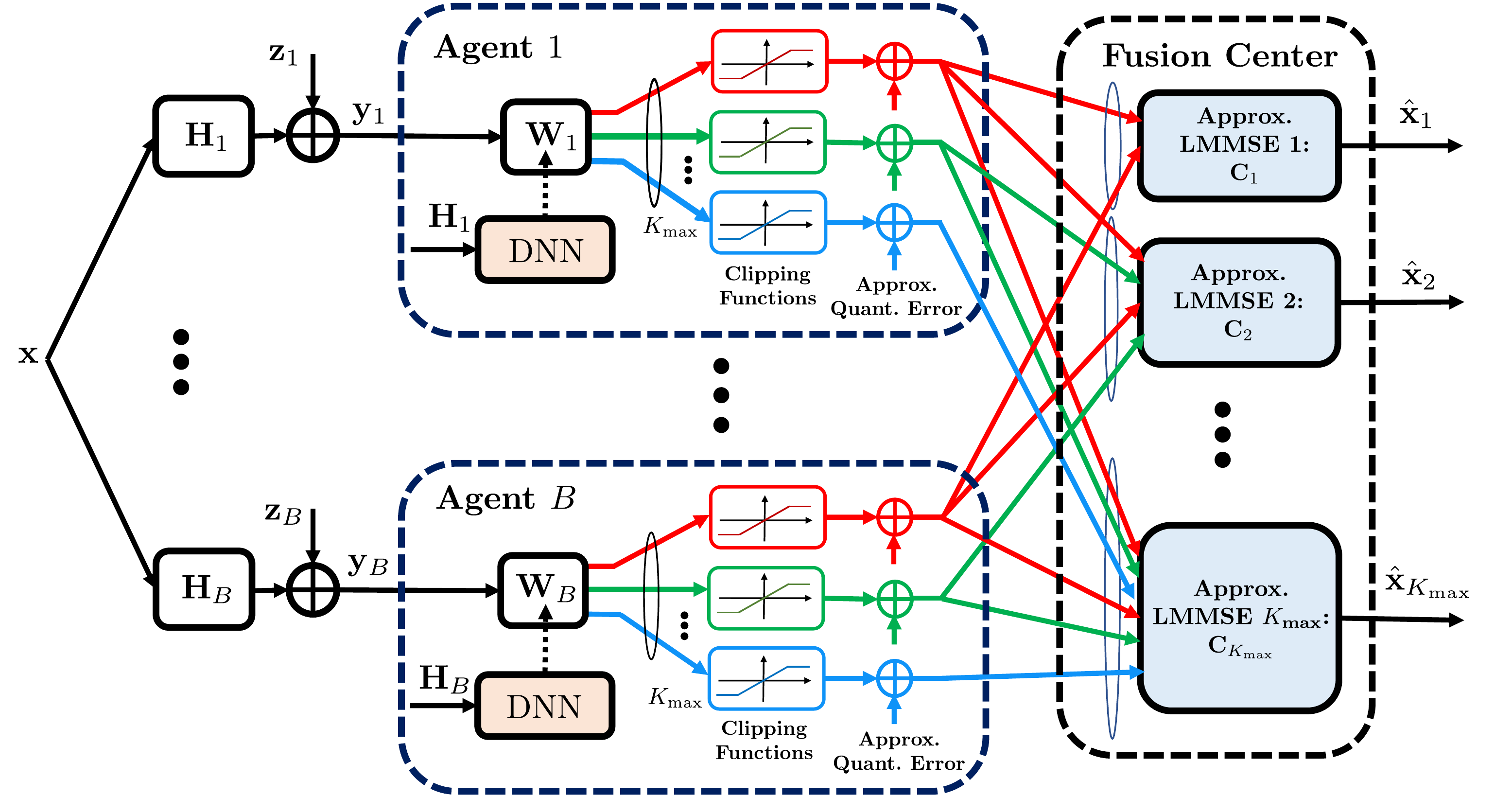}
     \caption{Block diagram of the proposed neural network architecture for designing an end-to-end distributed estimation system with progressive compression with maximum $K_\text{max}$ stages.}
     \label{fig:proposed}
\end{figure*} 

To the best of the authors' knowledge, among the conventional designs for distributed wireless networks only the PCA-based methods, e.g. \cite{Liang2015CRAN,Zhu2005EVD,TomLuo2007BCD2}, in which each agent utilizes the local CSI to design the compression matrices admit the progressive compression scheme in problem formulation \eqref{eq:problem_final}. More specifically, in these designs, the $k$-th row of the compression matrix at agent $i$ is set to be the eigenvector corresponding to the $k$-th largest eigenvalue of \changeN{$\bH_i \boldsymbol{\Sigma}_{\bx}\bH_i^\top + \sigma^2 \mathbf{I}$ 
where $\boldsymbol{\Sigma}_{\bx}$ is the covariance matrix of $\bx$ computed based on $f_\bx$.} In the rest of the paper, PCA is also referred to as the eigenvalue decomposition (EVD) based scheme.

The PCA or EVD-based scheme is however developed for the single-agent scenario. In this paper, we show that the performance of the PCA or EVD-based approach can be significantly improved if a deep learning framework is used to design the progressive compression matrices (based on local CSI) in a data-driven fashion.

\section{End-to-End Design of Progressive Distributed Compression Strategies}
\label{sec:proposed}

In this section, we present how to design progressive distributed compression strategies using a deep learning framework. Furthermore, we discuss how to train the proposed deep learning architecture to jointly optimize the compression and quantization schemes at the agents as well as the estimation scheme at the fusion center.

\subsection{DNN Architecture}

The design of distributed estimation strategies involves optimizing three main components: (i) compression matrix at each agent, (ii) quantization scheme at each agent, and (iii) the aggregation matrix at the fusion center. In this subsection, we explain how each of these components is modeled in the proposed deep learning framework shown in Fig.~\ref{fig:proposed}.    

\subsubsection{Compression Matrix Design}

As mentioned earlier, each agent needs to design its compression matrix $\mathbf{W}_i\in \mathbb{R}^{K_\text{max}\times N}$ based on the available local CSI matrix $\mathbf{H}_i$. To do so, agent $i$ employs an $L$-layer fully-connected DNN that maps the local CSI to the compression matrix as:
\begin{equation} 
\label{eq:dnn}
\widetilde{\bw}_i = \sigma_L^{(i)}\left(\bA_L^{(i)}  \sigma_{L-1}^{(i)}\left(\cdots \sigma_1^{(i)}\left(\bA_1^{(i)}\widetilde{\bh}_i + \bb_1^{(i)} \right)  \cdots \right) + \bb_L^{(i)}\right), 
\end{equation}
where $\widetilde{\bh}_i \triangleq \operatorname{vec}(\bH_i)$ and $\widetilde{\bw}_i \triangleq \operatorname{vec}(\bW_i)$ are respectively the vectorized representations of the observation matrix and the compression matrix, $\left\{\bA_\ell^{(i)}, \bb_\ell^{(i)} \right\}_{\ell=1}^L$ is the set of the trainable weights and biases in the fully-connected DNN of agent $i$, and $\sigma_\ell^{(i)}$ is the corresponding activation function of the $\ell$-th layer. Note that the activation functions at different layers, i.e., $\sigma_1^{(i)},\ldots,\sigma_{L}^{(i)}$, are design parameters of the network.

In the $k$-th stage of the progressive transmission scheme, agent $i$ utilizes the first $k$ rows of the compression matrix $\bW_i$ designed by the DNN according to \eqref{eq:dnn}. 

\subsubsection{Quantizer Design}

The next component that needs be designed at the $i$-th agent is the quantization scheme applied to the compressed signals $v_{i,k} = \mathbf{w}_{i,k}^\top \by_i$. In general, finding an optimal quantization scheme (i.e., discrete representation of a signal) using a deep learning framework is a challenging task. This is because the quantization operation which involves discretizing a continuous signal (almost) always leads to zero derivatives in the back-propagation step of the DNN training. As a result, gradients can never flow through a quantization operation, and consequently any trainable parameters before this operation cannot be trained with the conventional back-propagation method.

A commonly used trick in the machine learning literature to get around this issue is to employ a \emph{straight-through} estimator which approximates the quantization operation in the back-propagation step with a smooth differentiable function, e.g., \cite{hinton2012videos,bengio2013,chung2016}. However, training a DNN with straight-through approximation for quantization can be challenging \cite{Foad2019Asilomar}. In this paper, inspired by the way that the quantization error is modeled in the information theory literature \cite{GrayQuantization}, we propose an alternative novel method to train a quantization scheme within a deep learning framework. In particular, we propose to fix the quantization scheme in each dimension to a uniform quantizer \changeF{\cite{gallager2008principles}} with a designable dynamic range. If we assume that the dynamic range of the $k$-th dimension for the $i$-th agent is given, i.e., $q_{i,k}$, we then model the quantization error of a $Q$-bit uniform quantization as a uniformly distributed noise as follows:
\begin{equation}
    \tilde{v}_{i,k} = {v}_{i,k} + e_{i,k},
\end{equation}
where
\begin{equation}
    e_{i,k} \sim \mathcal{U}\left(\frac{-q_{i,k}}{2^Q},\frac{q_{i,k}}{2^Q} \right).
\end{equation}
We remark that this model for quantization error is only used in the DNN training phase, i.e., in the operational phase (i.e., test phase), we perform the obtained uniform quantization scheme. \changeF{Moreover, to enforce that the signal $v_{i,k}$ falls into the presumed dynamic range $[-q_{i,k}, q_{i,k}]$, we apply a clipping function $c(\cdot)$ as follows:
\begin{equation}
    c(v_{i,k}) = \begin{cases}
    -q_{i,k}, \quad \text{if} \quad v_{i,k}<-q_{i,k},\\
    v_{i,k}, \hspace{18pt} \text{if} \quad -q_{i,k}<=v_{i,k}<=q_{i,k},\\
    q_{i,k}, \hspace{18pt} \text{if} \quad q_{i,k}<v_{i,k}.
    \end{cases}
\end{equation}
}

Now, the remaining question is how to learn the dynamic range $q_{i,k}$. One
straightforward idea is to make $q_{i,k}$ as a trainable parameter in the deep
learning architecture. However, numerical results show that learning the dynamic range in this way is inefficient.  This is
because the distribution of the variable to be quantized (i.e., $v_{i,k} =
\mathbf{w}_{i,k}^\top \by_i$) changes during the training process, as the DNN
parameters that lead to the design of $\mathbf{w}_{i,k}$ (as in \eqref{eq:dnn})
change. 


This paper recognizes that the challenge of designing the dynamic range of a uniform quantizer is similar to the challenge of training a multi-layer DNN where the distribution of each layer's inputs changes during training, as the parameters of the previous layers change \cite{Ioffe2015}. In the multi-layer DNN training context, \cite{Ioffe2015} shows that the statistical information in training batch data can be directly exploited to tackle this so-called \emph{internal covariate shift} phenomenon. Motivated by the results of \cite{Ioffe2015}, we propose to use the statistics of the batch training data to obtain a reasonable design for the dynamic range of ${v}_{i,k}$. In particular, since the appropriate quantization dynamic range is typically closely related to the standard deviation of the parameter that needs be quantized, we propose to set the dynamic range as a scaled version of the standard deviation computed based on the batch data  $\mathcal{B} = \left\{ {v}_{i,k}^{(j)}
\right\}_{j=1}^{|\mathcal{B}|}$ as:
\begin{equation}
    q_{i,k} = s_{i,k} \hspace{2pt} \sigma_{{v}_{i,k}},
    \label{eq:qik}
\end{equation}
where 
\begin{subequations}
    \begin{align}
        \sigma_{{v}_{i,k}} &= \sqrt{\frac{1}{|\mathcal{B}|} \sum_{j=1}^{|\mathcal{B}|} \left({v}_{i,k}^{(j)} - \mu_{{v}_{i,k}}\right)^2},\\
        \mu_{{v}_{i,k}} &= \frac{1}{|\mathcal{B}|} \sum_{j=1}^{|\mathcal{B}|} {v}_{i,k}^{(j)},
    \end{align}
\end{subequations}
and $s_{i,k}$ is an appropriate scaling factor. For example, for a Gaussian distributed ${v}_{i,k}$, a good choice for the scaling factor is $s_{i,k} \in [3,4]$ to ensure that more than $99\%$ of the time the realization of ${v}_{i,k}$ is within the quantization dynamic range. But, to avoid setting the scaling factor based on any heuristic measures, we consider $s_{i,k}$ to be trainable in the proposed deep learning framework.

\begin{figure}[t]
     \centering
     \includegraphics[width=0.5\textwidth]{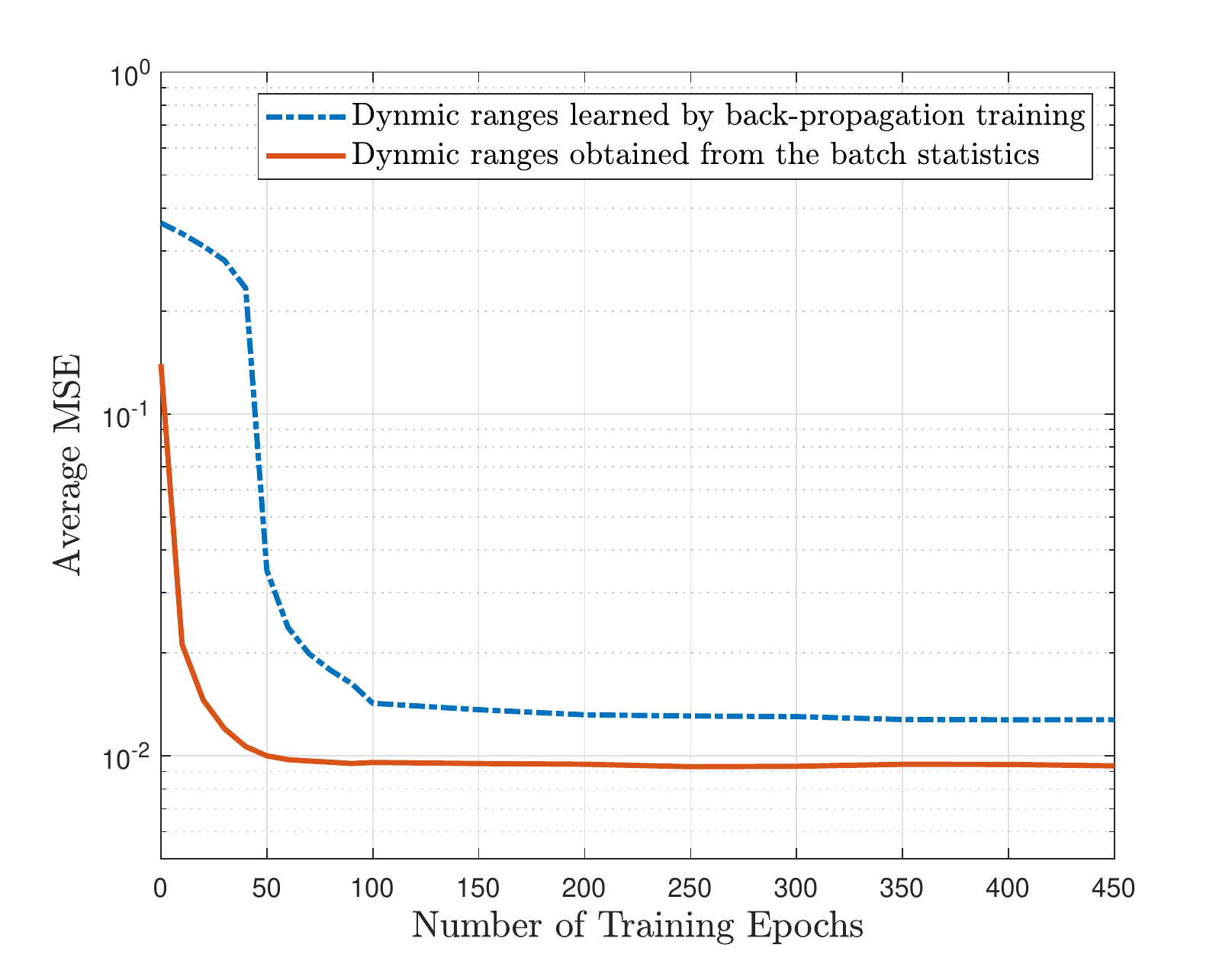}
     \caption{Average MSE versus the number of training epochs for two different quantization dynamic range designs in a distributed compression setup with $M=64$, $N=6$, $K=4$, $\bx \sim \mathcal{N}(\mathbf{0},\mathbf{I})$,  $\bH_i \sim \mathcal{N}(\mathbf{0},\mathbf{I})$, and $\operatorname{SNR} = 0$dB.}
     \label{fig:dynamic_Range}
\end{figure}

To illustrate the advantages of the proposed dynamic range design based on batch statistics as compared to the idea of training the dynamic ranges using back-propagation procedure, Fig.~\ref{fig:dynamic_Range} plots the average MSE against the number of training epochs for these two methods
in a distributed estimation setup with non-progressive compression where $M=64$, $N=6$, $K=4$, $\bx \sim \mathcal{N}(\mathbf{0},\mathbf{I})$,  $\bH_i \sim \mathcal{N}(\mathbf{0},\mathbf{I})$, and $\operatorname{SNR}\triangleq \log_{10}(\tfrac{1}{\sigma^2}) = 0$dB. It can be seen from Fig.~\ref{fig:dynamic_Range} that the proposed design utilizing the batch statistics is superior to the back-propagation training based design both from convergence and performance perspectives, i.e., the performance of the proposed design converges only after 50 epochs to a lower average MSE as compared to that of the back-propagation based design which converges after 100 epochs. 

\subsubsection{Progressive Linear Estimation Design}

Finally, we discuss how to set the linear estimator by designing the mapping function $\mathcal{G}(\cdot)$ at the fusion center. To do so, we can potentially model and learn the mapping function $\mathcal{G}(\cdot)$ by using another multi-layer DNN. However, through simulations, we observe that an approximate\footnote{We call the linear estimator in \eqref{eq:LMMSE} \changeN{the} approximate LMMSE since this is the LMMSE estimator when unquantized compressed signals are assumed to be available at the fusion center.} LMMSE estimator in which the effect of quantization error is ignored already leads to excellent performance for each of the $K_\text{max}$ stages of the progressive transmission, thus there is no need to use another DNN for this part.

To obtain the expression for such an approximate LMMSE estimator, suppose that $Q\to \infty$ so that we can ignore the quantization error. Under this assumption, the collection of the received compressed signals at the fusion center from all $B$ agents after the first $k$ stages of the progressive compression is in the form of $\breve{\bv}_{[k]} \triangleq [\tilde{\bv}_{1,[k]}^\top,\ldots,\tilde{\bv}_{B,[k]}^\top]^\top = \bU_{k} \bx + \tilde{\bz}_k$, where 
\begin{equation}
    \bU_{k} \triangleq {\begin{bmatrix}
      \bW_{1,[k]}\bH_1 \\[0.4em]   
      \vdots \\[0.4em]      
     \bW_{B,[k]}\bH_B
    \end{bmatrix}} \
\end{equation}
is the effective observation matrix after $k$ stages of progressive compression at the fusion center and $\tilde{\bz}_k \sim \mathcal{N}\left( \mathbf{0}, \boldsymbol{\Sigma}_{\tilde{\bz}_k} \right)$ with 
\begin{equation}
    \boldsymbol{\Sigma}_{\tilde{\bz}_k} \triangleq {\begin{bmatrix}
      \bW_{1,[k]}\bW_{1,[k]}^\top & \hspace{-7pt} \mathbf{0} & \hspace{-7pt}\ldots &\hspace{-7pt} \mathbf{0} \\[0.4em]   
      \mathbf{0} & \hspace{-7pt} \bW_{2,[k]}\bW_{2,[k]}^\top & \hspace{-7pt}\ldots & \hspace{-7pt}\mathbf{0} \\[0.4em]  
      \vdots & \hspace{-7pt} \vdots &\hspace{-7pt} \ddots &\hspace{-7pt} \vdots \\[0.4em]      
     \mathbf{0} & \hspace{-7pt} \mathbf{0} & \hspace{-7pt} \ldots & \hspace{-7pt}\bW_{B,[k]}\bW_{B,[k]}^\top
    \end{bmatrix}} \
\end{equation}
as the covariance matrix of the corresponding effective noise. Accordingly, the LMMSE estimation matrix is given by:
\begin{equation}
    \bC_k  = \boldsymbol{\Sigma}_{\bx} \bU_{k}^\top \left(\bU_{k}\boldsymbol{\Sigma}_{\bx}\bU_{k}^\top + \boldsymbol{\Sigma}_{\tilde{\bz}} \right)^{-1},
    \label{eq:LMMSE}
\end{equation}
where $\boldsymbol{\Sigma}_{\bx}$ is the covariance matrix of $\bx$ which can be computed based on the probability density function $f_\bx$. The excellent performance of this estimator is due to the fact that in the absence of quantization error, the LMMSE estimator in \eqref{eq:LMMSE} is the optimal linear estimator, and the performance degradation in the presence of the quantization error seems marginal.

The block diagram of the overall proposed neural network architecture that represents an end-to-end distributed wireless network with progressive transmission is illustrated in Fig.~\ref{fig:proposed}. In this neural network, the trainable parameters are the DNN parameters $\left\{\bA_\ell^{(i)}, \bb_\ell^{(i)} \right\}_{\ell=1}^L$ at the agents for designing the compression matrices and the scaling factors $\{\{s_{i,k}\}_{k=1}^{K_\text{max}}\}_{i=1}^{B}$ for determining the quantization dynamic ranges for $K_\text{max}$ transmission stages from the agents to the fusion center.

\subsection{DNN Training}

We now describe the training procedure of the DNN architecture in Fig.~\ref{fig:proposed}. To do so, we need to define the loss function. As explained earlier, the objective of interest in this paper is to minimize the aggregation of the distortion metrics for $K_{max}$ different estimators in the progressive transmission scheme, i.e., $\mathcal{D}\left(d_1,\ldots,d_{K_\text{max}}\right)$. Similar to \cite{Gunduz2019MLProgressive}, this paper uses the sum of the distortions as the aggregated cost function, i.e., $\sum_{k=1}^{K_\text{max}} d_k = \sum_{k=1}^{K_\text{max}} \| \hat{\bx}_k - \bx \|^2_2$. Accordingly, in the training procedure, we seek to design the trainable parameters by solving the following optimization problem:
\begin{equation}
    \min_{\{\Theta_i\}_{i=1}^B, \{\bs_i\}_{i=1}^{B}} \mathbb{E}_{\bH,\bx,\bz}\left[ \sum_{k=1}^{K_\text{max}}  \| \hat{\bx}_k - \bx \|^2_2 \right],
\end{equation}
where $\Theta_i \triangleq \left\{\bA_\ell^{(i)}, \bb_\ell^{(i)} \right\}_{\ell=1}^L$,  $\bs_i \triangleq \{s_{i,k}\}_{k=1}^{K_\text{max}}$, and the expectation is over the distribution of channels $\bH \triangleq [\bH_1^\top,\ldots,\bH_B^\top]^\top$, the distribution of the unknown signal $\bx$, and the distribution of the noise at the agents $\bz \triangleq [\bz_1^\top,\ldots,\bz_B^\top]^\top$. The parameter space consists of the compression matrix design of the agents and the corresponding quantization dynamic range design.

We assume certain distributions of the observation matrices $\bH$, signal of interest $\bx$, and the agent noise $\bz$ and accordingly generate a large data set consisting of the realizations of $\bH$, $\bx$, and $\bz$ for training purpose. The training problem  can then be efficiently tackled by employing a stochastic gradient descent (SGD) algorithm.

\subsection{Implementation Details}
\label{sec:imp_det}

We implement the proposed deep active learning framework on TensorFlow \cite{tensorflow2016} and Keras \cite{chollet2015} platforms and employ Adam optimizer \cite{adam2014} with a learning rate progressively decreasing from $10^{-4}$ to $10^{-5}$. For the agent-side DNNs \changeN{which map} the local CSI $\bH_i$ available at agent $i$ to the design of the compression matrix $\mathbf{W}_i$, $2$-layer neural networks with dense layers of widths $[2048,1024,MK_\text{max}]$ are employed. For faster convergence,
each dense layer is preceded by a batch normalization
layer \cite{Ioffe2015}. Further, we set the hidden layers' activation functions to the  hyperbolic tangent function, i.e., $\sigma_\ell^{(i)}(\cdot) =\operatorname{tanh}(\cdot), ~ \ell = 1,\ldots,L-1$. Since scaling the compression matrix does not affect the system performance, we consider a normalization activation function at the last layer of the form $\sigma_L^{(i)}(\cdot)=\tfrac{\cdot}{\|\cdot\|_2}$ to ensure that the entries of the matrix $\bW_i$ are not getting unboundedly large in the training phase. 
Moreover, in order to accelerate the training procedure, the trainable weights and biases of the hidden layers for different agents are tied together, i.e., $\bA_\ell^{(i)} = \bA_\ell$ and  $\bb_\ell^{(i)} = \bb_\ell,~ \forall \ell=1,\ldots,L-1$, while different dense layers are employed at the final layers to enable different agent DNNs to compress the received signals into different directions, even if their corresponding observation matrices are quite similar. Finally, in order to optimize the scaling factors $s_{i,k}$'s in the expression of the dynamic ranges $q_{i,k}$'s \eqref{eq:qik}, we define $s_{i,k}$'s as training variables in TensorFlow whose initial values are set to $4$.

To study the ultimate performance of the proposed deep learning approach, we assume that \changeF{the distributions of the channel $\bH$, the  source $\bx$, and the noise $\bz$ are known, so that we can generate as many training data samples as needed to fully train the proposed deep learning module.} We monitor the performance of the overall neural network during training by evaluating the empirical average of the loss function for an out-of-sample data set of size $10^5$ and keep the network model parameters that have produced the best validation-set performance so far. The training procedure is terminated when the performance of the validation set has not improved over several epochs.

\section{Numerical Results}
\label{sec:sims}
We now evaluate the performance of the proposed deep learning-based design for a distributed wireless network. We compare the performance of the proposed approach against several existing benchmarks. Before presenting the numerical results, we first briefly explain the considered baselines. 

\subsection{Baseline Strategies}

\subsubsection{Optimal estimation with no fronthaul bandwidth constraints} In this centralized scheme, we assume that there are no bandwidth constraints for the links between the agents and the fusion center. As a result, the full CSI ${\bH}$ and the entire observations $\by \triangleq [\by_1^\top,\ldots,\by_B^\top]^\top = \bH \bx + \bz$ can be made available at the fusion center, and the optimal linear estimate of $\bx$ can be obtained by applying the LMMSE as:
\begin{equation}
    \hat{\bx}^{\star}  = \boldsymbol{\Sigma}_{\bx} \bH^\top \left(\bH \boldsymbol{\Sigma}_{\bx}\bH^\top + \sigma^2 \mathbf{I} \right)^{-1}  \by,
    \label{eq:LMMSE_optimal_LB}
\end{equation}
The performance of the optimal estimate in \eqref{eq:LMMSE_optimal_LB} provides a MSE lower bound for all other methods.

\subsubsection{Block coordinate descent (BCD) algorithm \cite{Roy2008BCD1,TomLuo2007BCD2}} In this non-progressive global CSI-based approach, at the beginning of each coherence interval, the agents send the CSI information to the fusion center. Under the assumption that $Q\to\infty$, the fusion center then applies the block coordinate descent algorithm to iteratively update the design of each compression matrix $\bW_i$ and the aggregation matrix $\bC$. Once the algorithm converges, the fusion center sends back the design of $\bW_i$ to agent $i$. To have a fair comparison with the proposed approach, we set the quantization dynamic range \changeN{of each dimension to be a} scaled version of the corresponding standard deviation with the near-optimal scaling factors chosen by exhaustive search.

\begin{figure}[t]
     \centering
     \includegraphics[width=0.5\textwidth]{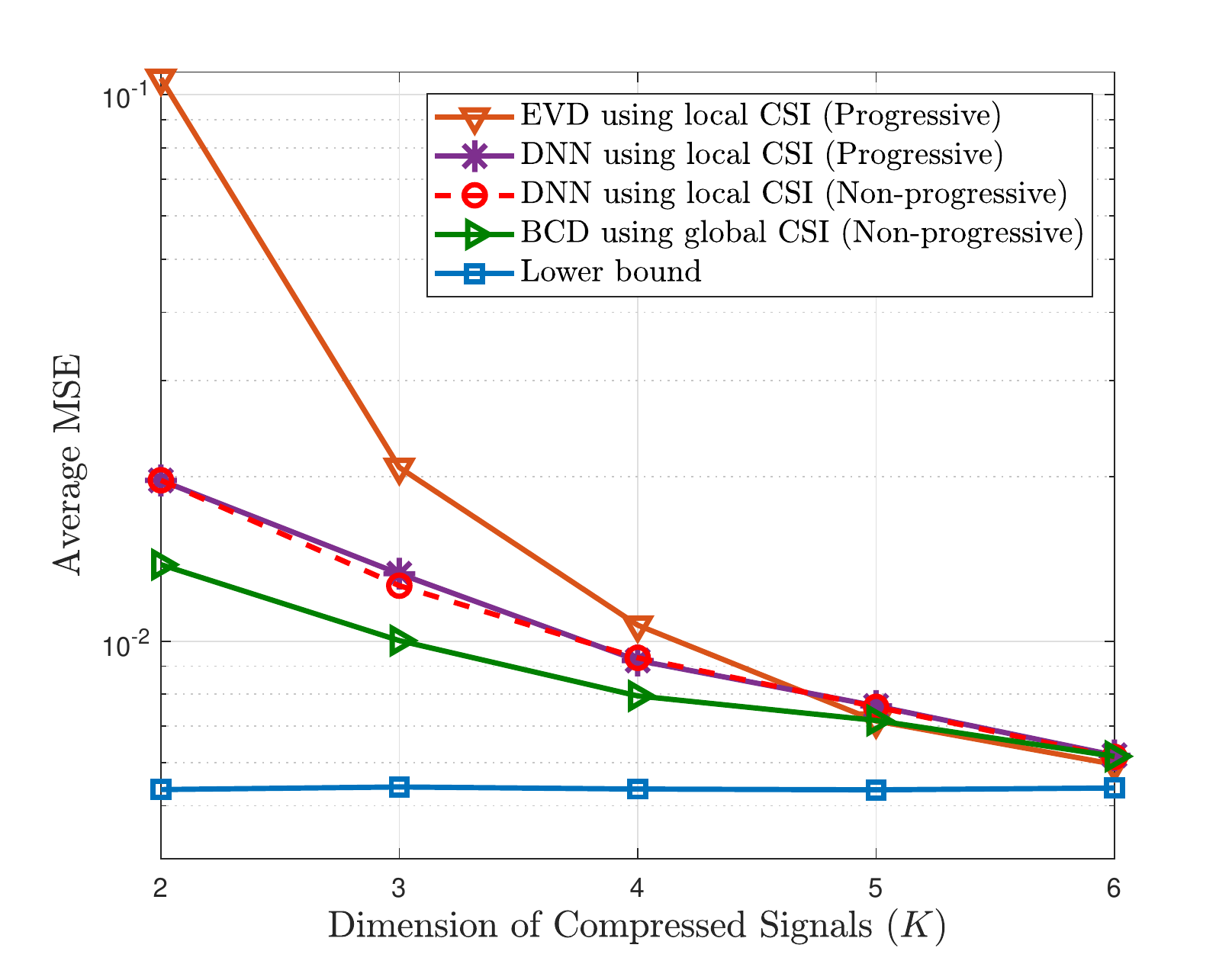}
     \caption{Average MSE versus the number of dimensions $K$ that each agent transmits to the fusion center in a distributed compression setup with $M=64$, $N=6$, $B=3$, $Q=6$~bits per dimension, uncorrelated source $\bx \sim \mathcal{N}(\mathbf{0},\mathbf{I})$,  $\bH_i \sim \mathcal{N}(\mathbf{0},\mathbf{I})$, and $\operatorname{SNR} = 0$dB. For the algorithms with progressive transmission, we set $K_\text{max}=6$.}
     \label{fig:uncorrelated_sweepK}
\end{figure}

\subsubsection{EVD based design with local CSI \cite{TomLuo2007BCD2,Liang2015CRAN}} In this baseline method, each agent designs its compression matrix by PCA in which the $k$-th row of the compression matrix at agent $i$ is set to be the eigenvector corresponding to the $k$-th largest eigenvalue of the covariance matrix of the received signal, i.e., \changeN{$\bH_i\boldsymbol{\Sigma}_\bx \bH_i^\top + \sigma^2 \bI$}. It can be seen that such a compression matrix design naturally leads to a progressive compression scheme. Given the design of the compression matrices, we then use the LMMSE estimator in \eqref{eq:LMMSE}. Finally, for the quantization dynamic range design, we follow the exhaustive search method explained in the BCD baseline. 

\subsection{Proposed DNN Approach}

In the first experiment, we evaluate the performance of the proposed approach as a function of the number of compression dimensions $K$. To do so, we consider a distributed estimation setup with $N=6$, $M=64$, $B=3$, and the number of quantization bits per dimension is $Q=6$. Fig.~\ref{fig:uncorrelated_sweepK} plots the average MSE over $10^4$ channel realizations for different methods against $K$. It can be seen from Fig.~\ref{fig:uncorrelated_sweepK} that the proposed deep learning-based method can achieve significantly smaller MSE as compared to the EVD-based baseline when the compression dimension $K$ is relatively small. 
This indicates that the proposed approach is well suited for scenarios in which the bandwidths of the fronthaul links are limited. 
The gain comes from the fact that the EVD-based scheme is a heuristic and is not optimized for distributed compression.

Moreover, Fig.~\ref{fig:uncorrelated_sweepK} shows that the progressive DNN architecture can achieve almost identical performance to the non-progressive DNN counterpart, verifying the effectiveness of the proposed DNN framework to design progressive compression protocols. This observation is aligned with the results in \cite{4475358} that a Gaussian source under the MSE distortion criterion is successively refinable for the Gaussian CEO problem. 

\begin{figure}[t]
     \centering
     \includegraphics[width=0.5\textwidth]{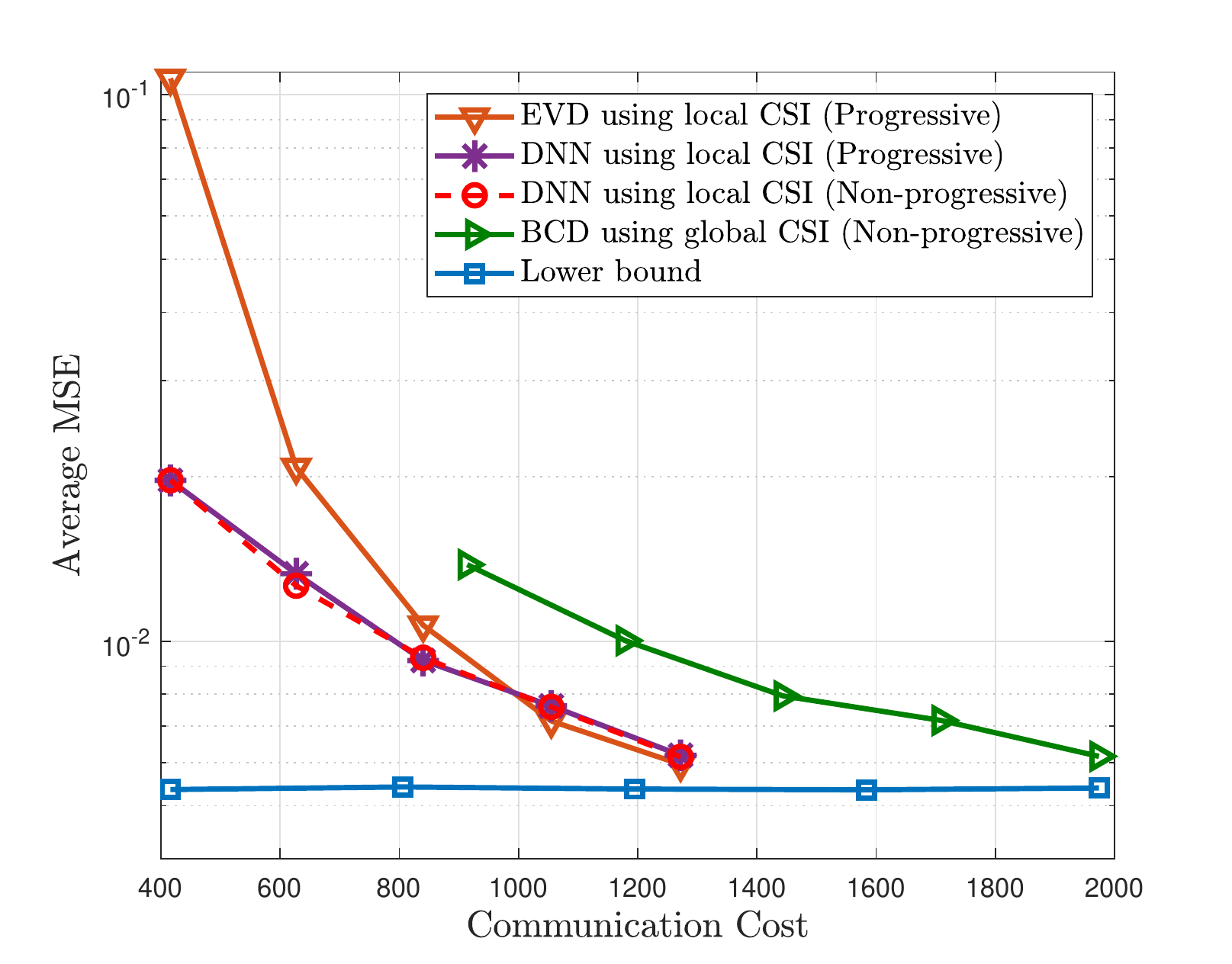}
     \caption{Average MSE versus communication cost (i.e., $C_\text{global}$ or $C_\text{local}$) in a distributed compression setup with $M=64$, $N=6$, $B=3$, $T=200$, $K\in\{2,\ldots,6\}$, $Q=6$~bits per dimension, uncorrelated source $\bx \sim \mathcal{N}(\mathbf{0},\mathbf{I})$,  $\bH_i \sim \mathcal{N}(\mathbf{0},\mathbf{I})$, and $\operatorname{SNR} = 0$dB. For the algorithms with progressive compression, we set $K_\text{max}=6$.}
     \label{fig:uncorrelated_sweepTX}
\end{figure}

As the number of dimensions increases, it is clear that the performances of all methods converge to the optimal lower bound. Note that at $K=6$, the small gap between the performance of all methods and the MSE lower bound is entirely due to the quantization error since we have $K=N$. The fact that the gap is small indicates that $Q=6$ is sufficient. We can further observe that the BCD algorithm outperforms the proposed method for a fixed $K$. However, we remark that the BCD algorithm designs the compression matrices at the fusion center based on the global CSI, which can lead to a larger communication overhead.

\subsection{Global vs. Local CSI}

To see the advantages of designing the compression matrices based only on the local CSI available at the agents, we compare the performances of different methods accounting for the overhead of communicating the CSI. Fig.~\ref{fig:uncorrelated_sweepTX} plots the average MSE achieved by different methods against the overall communication cost, i.e., $C_\text{global}$ given in \eqref{eq:C_global} for global CSI-based methods and $C_\text{local}$ given in \eqref{eq:C_local} for local CSI-based methods, for a distributed estimation setup with the same parameters as in the previous experiment, but with a channel coherence time of $T=200$ (i.e., the channels stay constant while the source is i.i.d.~and needs to be estimated over 200 time slots). It can be seen that the BCD algorithm utilizing the global CSI needs a much \changeN{larger} amount of overall communications to achieve a comparable MSE level as other methods which use the local CSI. This is because of the significant communication overhead required for sending the high-dimensional channel matrices from the agents to the fusion center in the global CSI-based designs. In contrast, the proposed approach only uses local CSI to design the compression scheme at the agents and requires only the lower-dimensional effective observation matrices for designing the aggregation function at the fusion center. In this way, the signaling overhead for achieving the same MSE value is significantly reduced. 

\begin{figure}[t]
     \centering
     \includegraphics[width=0.5\textwidth]{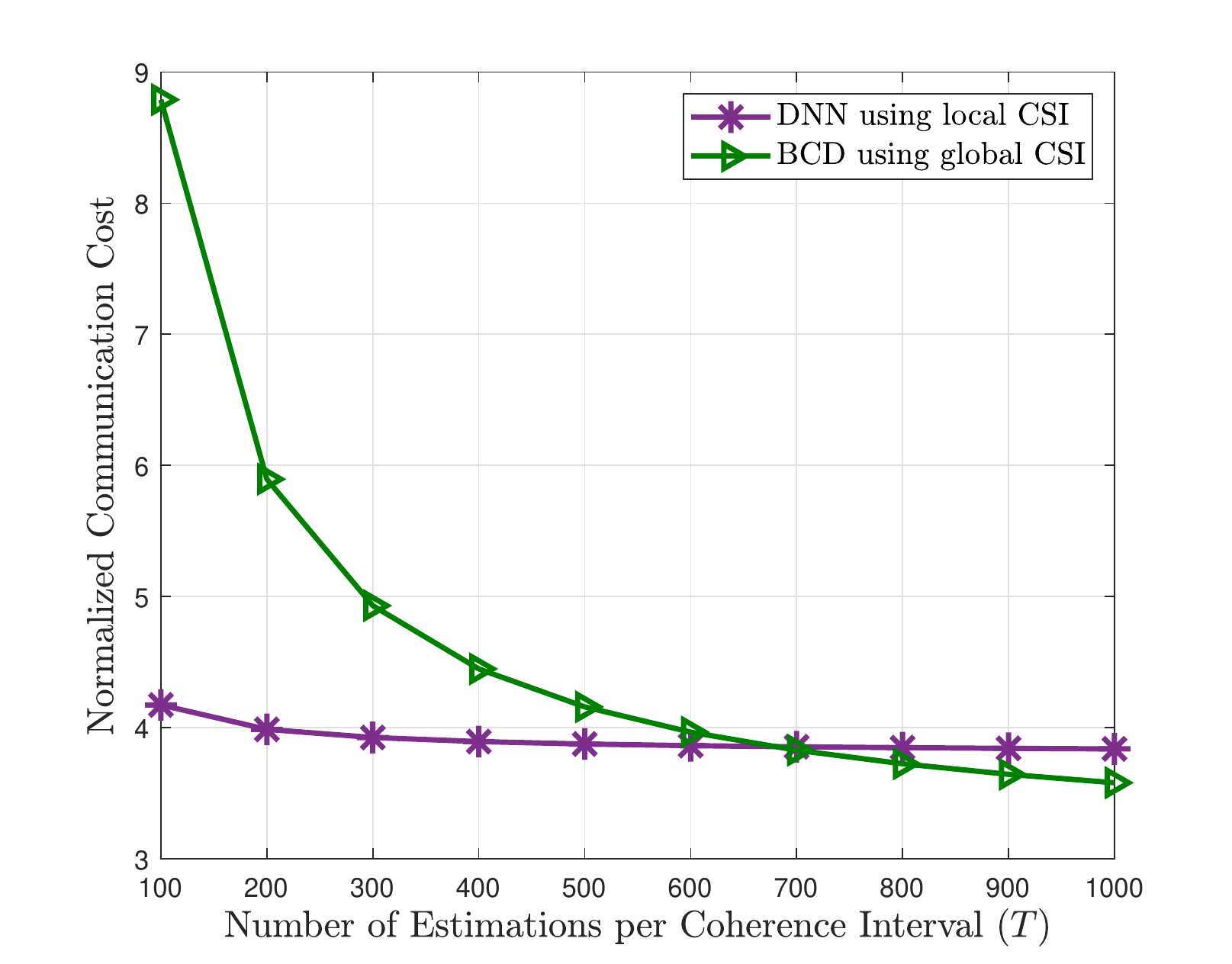}
     \caption{Normalized communication cost (i.e., $C_\text{local}/T$ or $C_\text{global}/T$)  required for the proposed DNN and the BCD algorithm to achieve the MSE distortion of $10^{-2}$ versus the number of estimations supported within a single coherence interval, i.e., $T$. The system parameters are the same as in Fig.~\ref{fig:uncorrelated_sweepK}.}
     \label{fig:uncorrelated_LocalvsGlobal}
\end{figure}

The advantage of local CSI methods depends on the number of estimations within the coherence time, $T$. 
Next, we seek to show the effect of $T$ on the overall communication cost comparison between the proposed approach and the BCD algorithm. Suppose that we require the MSE distortion level to be $10^{-2}$. Accordingly, from Fig.~\ref{fig:uncorrelated_sweepK}, we find the value of $K$ for each method that achieves the MSE distortion of $10^{-2}$. For such values of $K$, Fig.~\ref{fig:uncorrelated_LocalvsGlobal} plots the normalized communication cost for the proposed approach and the BCD algorithm, i.e., \changeN{${C_\text{local}}/{T}$ and ${C_\text{global}}/{T}$}, against the number of estimations within each coherence interval, $T$. As it can be seen from Fig.~\ref{fig:uncorrelated_LocalvsGlobal}, the proposed approach has a smaller normalized signaling overhead if $T<700$. This implies that the proposed local CSI-based approach is suitable whenever the coherence time is smaller than several hundreds of time slots, while the global CSI-based methods such as the BCD algorithm are preferred for the truly static channel scenario where the coherence time is very large.   

\begin{figure}[t]
     \centering
     \includegraphics[width=0.5\textwidth]{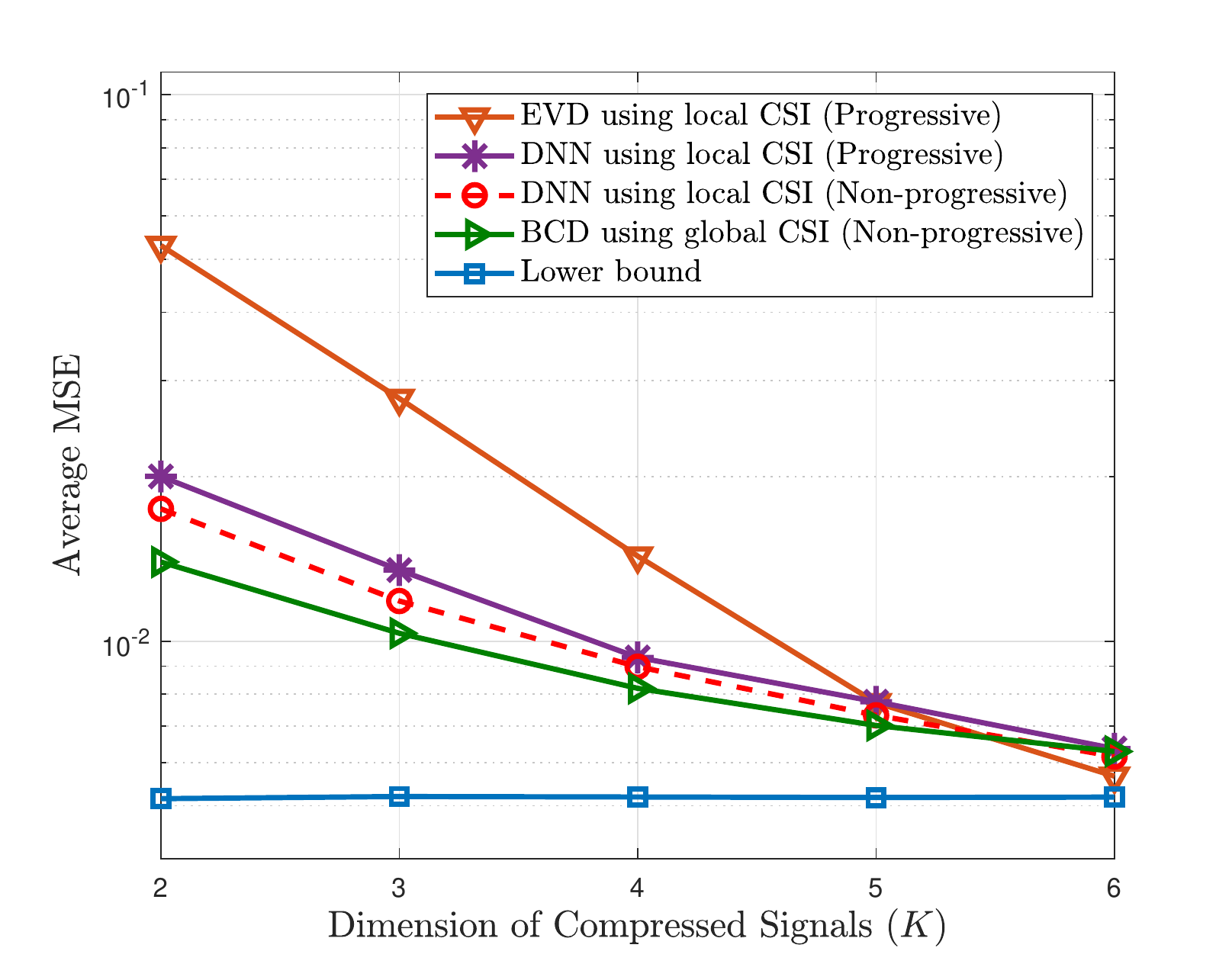}
     \caption{Average MSE versus the number of dimensions $K$ that each agent transmits to the fusion center in a distributed compression setup with $M=64$, $N=6$, $B=3$, $Q=6$~bits per dimension, correlated source $\bx \sim \mathcal{N}(\mathbf{0},\boldsymbol{\Sigma_x})$ where $\boldsymbol{\Sigma}_x$ is given in \eqref{eq:Sigma}, $\rho=0.9$, $\bH_i \sim \mathcal{N}(\mathbf{0},\mathbf{I})$, and $\operatorname{SNR} = 0$dB. For the algorithms with progressive transmission, we set $K_\text{max}=6$.}
     \label{fig:correlated_sweepK}
\end{figure}

\subsection{Correlated Source}

\begin{figure}[t]
     \centering
     \includegraphics[width=0.49\textwidth]{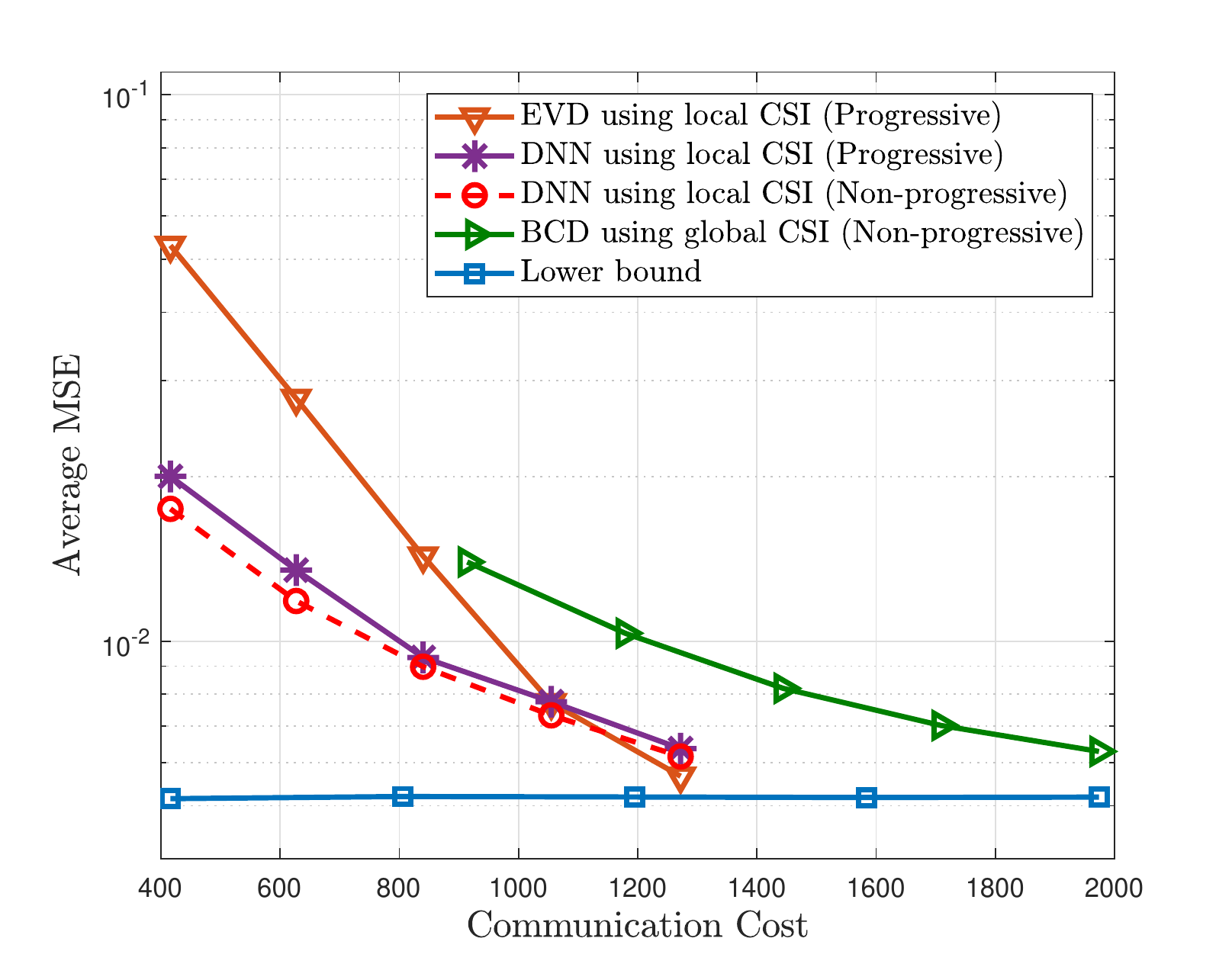}
     \caption{Average MSE versus communication cost (i.e., $C_\text{global}$ or $C_\text{local}$) in a distributed compression setup with $M=64$, $N=6$, $B=3$, $T=200$, , $K\in\{2,\ldots,6\}$, $Q=6$~bits per dimension, correlated source $\bx \sim \mathcal{N}(\mathbf{0},\boldsymbol{\Sigma_x})$ where $\boldsymbol{\Sigma}_x$ is given in \eqref{eq:Sigma}, $\rho=0.9$,  $\bH_i \sim \mathcal{N}(\mathbf{0},\mathbf{I})$, and $\operatorname{SNR} = 0$dB. For the algorithms with progressive transmission, we set $K_\text{max}=6$.}
     \label{fig:correlated_sweepTX}
\end{figure}

Finally, we evaluate the performance of the proposed method for the case when the source signal $\mathbf{x}$ is correlated, i.e., $\bx \sim \mathcal{N}(\mathbf{0},\boldsymbol{\Sigma_\bx})$ where the element in the $i$-th row and the $j$-th column of the covariance matrix is given by:
\begin{equation}
    \boldsymbol{\Sigma}_\bx(i,j) = \rho^{|i-j|},
    \label{eq:Sigma}
\end{equation}
and $\rho=0.9$. The other system parameters are the same as in Fig.~\ref{fig:uncorrelated_sweepK}. From Fig.~\ref{fig:correlated_sweepK}, we observe that the general behavior of different methods is similar to that of the uncorrelated source scenario in Fig. \ref{fig:uncorrelated_sweepK}. In particular, 
it can be seen that the performance gap between the progressive and the non-progressive compression schemes designed by the proposed DNN is still marginal. Further, the proposed deep learning approaches achieve much better performances as compared to the EVD-based approach when $K$ is small. It is also observed that the BCD algorithm which utilizes the global CSI can slightly outperform the proposed DNN which only uses the local CSI. However, when we take the overall required signaling into account, the performance of the proposed DNN is superior to the BCD for practical values of $T$, e.g., see Fig.~\ref{fig:correlated_sweepTX} in which $T=200$.

\section{Conclusion}
\label{sec:conclusion}

This paper proposes a deep learning framework to design the progressive dimension reduction matrix and quantization scheme at each agent for a distributed estimation task in a wireless network with fronthaul bandwidth constraints between the agents and the fusion center. In particular, the high-dimensional observations at each agent are compressed progressively using a linear matrix designed by the proposed deep learning framework, then quantized and transmitted to the fusion center for source reconstruction. Key features of the proposed learning framework are that it only requires local CSI at each agent and that compression can be done progressively. To tackle the challenge of training a DNN with quantization operations in the hidden layers, this paper proposes the novel idea of modeling the uniform quantization scheme by uniformly distributed quantization error inside the DNN and setting the quantization dynamic range based on the statistical information in the training data. Numerical results show that the proposed deep learning framework using local CSI requires significantly less communication cost to achieve the same MSE distortion as compared to the existing BCD algorithm that needs global CSI.

\changeF{We remark that one of the crucial aspects of employing machine learning methods in practice is that how generalizable their performance is when there is a mismatch between training and testing sample distributions. For example in the problem setting considered in this paper, it is interesting to investigate the performance of the proposed approach if the channel distribution in the training set is different from the testing phase. While our preliminary results show that the trained DNN for i.i.d.\ Gaussian channel model can still achieve excellent performance in the scenarios with the correlated Gaussian channels, a complete investigation of how to ensure generalizability for different settings is an interesting future direction.}
\bibliographystyle{IEEEtran}
\bibliography{IEEEabrv,referenceF}

\end{document}